\documentstyle{article}

\author{Zhen Wang\thanks{The author is thankful for Prof. Lin Xie for 
beneficial discussion. }\\
Physics Department, LiaoNing Normal University,\\
DaLian, PostCode 116029, P. R. China}
\title{What is in One: Uncertainty Quantum and Continuum Hypothesis
}
\input tcilatex

\QQQ{Description}{
}

\QQQ{Subject}{
02.10.By   03.65.Bz
}

\QQQ{Keywords}{
quantum, infinite, Continuum Hypothesis
}

\QQQ{Typist}{
}

\QQQ{Created}{
}

\QQQ{Language}{
American English
}

\begin{document}

\maketitle
\begin{abstract}
The concept of measurement is discussed. It is argued that counting process
in mathematics is also measurement which requires a basic unit. The idea of
scale is put forward. The basic unit itself, which are composed of the
infinitesimal of uncertainty quantum, can be regarded as infinite in another
scale. Thus infinite, infinitesimal and integer ''1'' are unified. It is
proposed that multiplication changes to summation when it is transformed to
a larger scale. The Continuum Hypothesis is proved to be correct after a
scale transformation.
\end{abstract}

In the history of physics and mathematics, it is often seen that progress in
one area depends on the progress in the other area, and difficulty in this
area connects subtly with the difficulty in that area. From the perspective
of physics, infinite in physics is a source of uncertainty and instability
which has been haunting for a long time. Though many mathematical techniques
have been developed in theory to deal with it, no further understanding of
the concept has been got since it was first introduced. This is mainly
because there has not been a breakthrough in the understanding of
measurement in physics. Though this may sound a little bit strange, it can
be seen when we think about the mathematical genesis of infinite. When
Cantor introduced the family of infinite, he got it by continuous counting
of integers. Obviously the counting is a measuring process with '' 1 '' to
be the basic unit. Thus if we hope to get deeper understanding about the
concept of infinite, we must have an inspection to the concept of
measurement in physics.

As is well known, the concept of measurement in quantum physics has been one
of the most controversial point in modern physics[1]. We believe that
progress for this problem relies on an overall and integrated understanding
of the concept of measurement in both quantum physics and classical physics.
One common feature of measurements in the two realms is: there is always a
smallest unit for any measurement. This is most significant through out all
our research[2,3]. Logically, no quantity can be made up of zeros. Thus for
a quantity to be meaningful, such smallest unit must exist. We give it the
name uncertainty quantum, denoted with $q$ . Now we shall see how these
uncertainty quanta make up a quantity.

Obviously, the uncertainty quantum can not be identified, though it can be
enumerated. We {\em can not} know its size, inner structure and other
information. If we could, it would not be the smallest unit, and then by
definition, it would not be the uncertainty quantum. This does not mean $q$
has no inner structure. Rather, it is too small at {\bf present scale}. If
we admit {\bf real infinite} in the following sense, $q$ can be connected
with $\omega $ , which represent enumerable infinite in mathematics. We know
the $\omega $ is derived in counting process:

$$
1,2,\cdots \cdots ,\,\omega ,\,\omega +1\cdots \cdots ,\,2\omega ,\,\,\cdots
\cdots ,\,\omega ^2,\,\cdots \cdots ,\,\omega ^\omega ,\cdots \cdots
\;\;\;\;\;\;(1) 
$$
We understand the {\bf real infinite} in such a sense that the infinite
becomes the new unit, thus makes us come to another scale. Since logic
requires the uncertainty quantum must exist, the infinite, which represent a
quantity in another scale, must exist for the same reason. No infinite in
reality is not relative. But just as the uncertainty quantum can not be
identified, it can also be understood that real infinite can not be
identified in present scale, and in a new scale it may be just a new
quantity. In this sense, any quantity may be expressed as 
$$
V=\,\omega q\;\;\;\;\;\;\;\;\;\;(2) 
$$
where $V$ may be any quantity. An infinite quantity arises when the
uncertainty quantum is taken to be the {\bf present unit} ''1 '', i.e., the
present unit '' 1 '' becomes infinitesimal so that there is no difference
between 1 and $n\,\;(n\prec \omega )$ . If we adopt suitable unit, (2) can
be expressed as:%
$$
q=\frac 1\omega \;\;\;\;\;\;\;\;(3)\; 
$$
(3) tells us how the uncertainty quantum connects with the enumerable
infinite, and together they make up a quantity. In such a sense, quantities
lose their absolute meaning, and present scale is no longer special. The
occurrence of infinite and infinitesimal just mean a scale-shift. Therefore,
the uncertainty quantum itself must also be regarded as a quantity composed
of uncertainty quantum in a smaller scale. Thus we can add a symmetric part
to (1):%
$$
\cdots \cdots ,\frac 1{\omega ^\omega }\,,\cdots \cdots ,\frac 1{\omega
^2}\,,\,\,\cdots \cdots ,\frac 1{2\omega }\,,\,\cdots \cdots ,\,\frac
1\omega \,,\cdots \cdots ,\,\frac
12\,,\,1\,\,\,\,\,\,\,\,\,\,\,\,\,\,\,\,\,\,\,\;\;\,\,\,(4) 
$$

So when infinite is involved, the basic unit '' 1 '' in the ordinal series
(1) can be regarded as the uncertainty quantum in this case. From the
perspective of mathematics, all terms after $\omega $ in (1) have the same
weight, the cardinal $\aleph _0$ [4], and the unit '' $1$ '' now has the
following property, which is the essential feature of uncertainty quantum:%
$$
\aleph _0\,+1=\aleph _0\;\;\;\;\;\;\;\;\;\;\;\;\;\;\;\;\;\;(5) 
$$
where $\aleph _0$ is cardinal number of $\omega $ . In the same way in the
series (4), all terms have the same uncertainty quantum $q$ . It is well
known that in the enumerating process, no new cardinals are generated. If we
define a new cardinal $\aleph _{-1}$ which corresponds to the uncertainty
quantum $q=\frac 1\omega $ , we can clearly figure out the scope of present
scale: from $\aleph _{-1}$ to $\aleph _0$ . Though we can describe a
quantity like $\sqrt{2}$ , we never use one with infinite decimal numbers in
measurement or calculation. Any quantity in reality must be within present
scale if it is to be clearly figured out.

Therefore any quantity outside present scale must be unenumerable, like the
cardinals of Cantor[5] which are bigger than $\aleph _0$ . When one deals
with structure beyond infinite or within uncertainty quantum, new cardinals
are introduced. If we add the symmetric part of cardinals in the other side,
we can write out the series:%
$$
\cdots \cdots ,\aleph _{-\alpha },\cdots \cdots ,\aleph _{-2}\,,\,\,\aleph
_{-1},\cdots \cdots 1,2,\,\cdots \cdots ,\aleph _0\,,\aleph _1,\cdots \cdots
,\,\aleph _\alpha ,\,\cdots \cdots \,\,\,\,\;\;\;\;\;(6) 
$$
in which each cardinal is the uncertainty quantum for the next bigger
cardinal.

We know these infinite cardinals have many miraculous properties, which
provide clues for the rules of scale-shift. Here we put forward a
proposition: {\em Summation would change to multiplication when it gets into
next bigger scale from present scale. }In fact, it is easily verified that%
$$
\begin{array}{c}
n_1n_2\aleph _0=n_1+n_2+\aleph _0=\aleph _0,\;\;\;n_1,n_2\leq \aleph
_0\;\;\;\;\;\;\;\;\;\,\,\;\;(7) \\ 
\aleph _\alpha +\aleph _\beta =\aleph _\alpha \cdot \aleph _\beta =\max
(\aleph _\alpha \,,\aleph _\beta )\,
\end{array}
$$
That means the infinite cardinal $\aleph _0$ has become a new basic unit for
variables in the new scale, while the old basic unit '' 1 '' has become the
uncertainty quantum.

This proposition sheds new light to the famous question Continuum Hypothesis
[5], which presumed that the weight of the real set $2^{\aleph _0}$ is the
first unenumerable infinite cardinal $\aleph _1$, i.e., 
$$
2^{\aleph _0}=\aleph _1\;\;\;\;\;\;\;\;\;(8) 
$$
This has to be transformed into present scale if it is to be understood.
According to the proposition, the right side of (8) will change to $\aleph _0
$ when it is transformed from next scale to present scale, the left side
will change to%
$$
2^{\aleph _0}=2\cdot 2\cdot 2\cdots \cdots \rightarrow 2+2+2+\cdots \cdots
=\aleph _0\,\,\,\,\;\;\;\;\;\;\;(9)\, 
$$
Therefore if we admit that arithmetical rules are not special for present
scale and can be extended to neighboring scale, the Continuum Hypothesis is
correct. Such proof may not be strict enough from the perspective of
mathematics, but physics picture helps us get new insight into the problem.
Actually, many mathematicians have already concluded[5] that the difficulty
of Continuum Hypothesis may not be mathematical in essence. What has
hindered mathematicians from the time of Cantor may be that the basic unit
'' 1 '' also has inner structure, and it is infinite itself from another
scale. Thus this problem connects profoundly with the concept of measurement
in philosophy as well as in physics. In a general theoretical framework[2]
this inner structure within the uncertainty quantum is shown to be connected
with the structure beyond infinite. We know on the real axis the number of
points within $(0,1)$ is equal to that of all points on the axis. This has
been taken to be a characteristic feature of a infinite set. In our
theory[2] this miraculous feature comes from breaking down of a symmetry in
counting or measurement, which gets realized in the unfurling of the
uncertainty quantum in time or space.

{\bf REFERENCES:}

[1] M. Jammer, {\it The Philosophy of Quantum Mechanics} (John Wiley, New
York 1974)

[2] Zhen Wang, quant-ph/9806071

[3] Zhen Wang, quant-ph/9804070

[4] J. W. Zhang and X. S. Wang, {\it Continuum Hypothesis} (LiaoNing
Education Press, 1988)

[5] P. J. Cohen, {\it Set Theory and Continuum Hypothesis} (W. A. Benjamin
Inc., New York 1966)

\end{document}